# Why a testing career is not the first choices of engineers?

Pradeep Kashinath Waychal, NMIMS University, Pune, India
Luiz Fernando Capretz, Western University, London, Ontario, Canada

## Introduction

Many software disasters – such as Ariane-5 [1] and the air traffic control system in LA airport [2] – have occurred in software product development. In fact, many others are happening as we write. The US-based National Institute of Standards and Technology (NIST) found, in its 2002 study, that the country is losing $59.5 billion each year due to software errors [3]. Charette [4] argues that we waste billions of dollars annually on entirely preventable mistakes in software development. Micro Focus [5] report points out that the effects of poor testing stretch beyond the back office; they also affect the boardroom and even the brand name. As software systems are becoming larger, more complex, and dependent on many third-party software components, the chances of their failure are increasing further. This calls for intense efforts to improve the quality of testing in the software development process.

The improvement can result from initiatives in process, technology, and people dimensions. Researchers and practitioners have paid adequate attention to the process [6] [7] and technology dimensions [8]; [9] Even though there are reports about the inadequacies of testing professionals and their skills [5], only a few studies tackle the problem [10]. The trait of paying less attention to the people dimension is observed throughout software engineering – the parent discipline of testing. [11]; [12]. Glass et al. [13] have studied 369 papers in six leading journals and discovered that software engineering research is fundamentally about technical and computing issues and seldom about behavioral issues. So it seems imperative that we explore the people dimension carefully, even though it is new to software engineering (and testing) researchers.

Human and social aspects play a significant role in software testing practices [14]. In an academic setting, attention to human factors in software testing have been preached by [15] and [16]. In a real-world environment, Shah & Harrold [17] found that software engineers with a positive attitude towards software testing can significantly influence those who have a negative attitute.

We can start with the basic problem that plagues the testing profession – the shortage of talent [5]. There are instituitions that teach and certify software testers, and business organizations that have developed career paths for testers. However, the main questions, is testing the first choice of software engineering students? If it is not, what are the reasons? Can the reasons be tackled in any way to change attitudes toward testing?

To answer these questions, we carried out a survey of junior students of a reputable computer engineering program, just before their internship and placement season. We asked them to list the pros and cons of a testing career and if they would choose one. We also carried out a survey of testing professionals, asking them to give the pros and cons of their profession. After analysing the reasons, we are proposing solutions to bring in changes to attract more students to testing careers. This would, in turn, help to improve the quality of the testing effort and of the software end-products. The next section covers the research design process and includes discussion and conclusion.

**Research Design**

Our study analyzed the reasons for computer engineering graduates not choosing testing careers. We asked a sample of students to provide pros and cons about the career. We compared the pros and cons from students with those provided by test professionals to propose possible remedial measures. The overall research design is outlined in Figure 1.

While software engineering is delivering unprecedented performance-to-cost ratio, it is also facing tough questions about glaring failures that have caused the loss of billions of dollars and even of human lives. Software architects, designers, and developers have been trying their best to reduce defects. They require the help of testers at various developmental stages to uncover defects before releasing their products to the end-users. Uncovering defects is the job of testers; however, very few bright individuals voluntarily choose testing careers. That has robbed the industry of opportunities for good testing and delivery of good quality products. To change this situation, it is necessary to analyze the reasons for such apathy towards testing careers.

Our research uses views on the pros and cons of testing careers from junior students of a computer-engineering department and working test professionals from the industry and is descriptive, diagnostic, cross-sectional, and field-setting. Descriptive research describes the characteristics of a population being studied and does not explore the reasons for those characteristics. Diagnostic research studies determine the frequency with which something occurs or its association with something else. We did not study the event over time but at a particular cross section, making the study cross-sectional. Our research covers real life situations and, therefore, is a field-setting study. We used a qualitative method by asking open-ended responses to the Pros and Cons of the testing career.

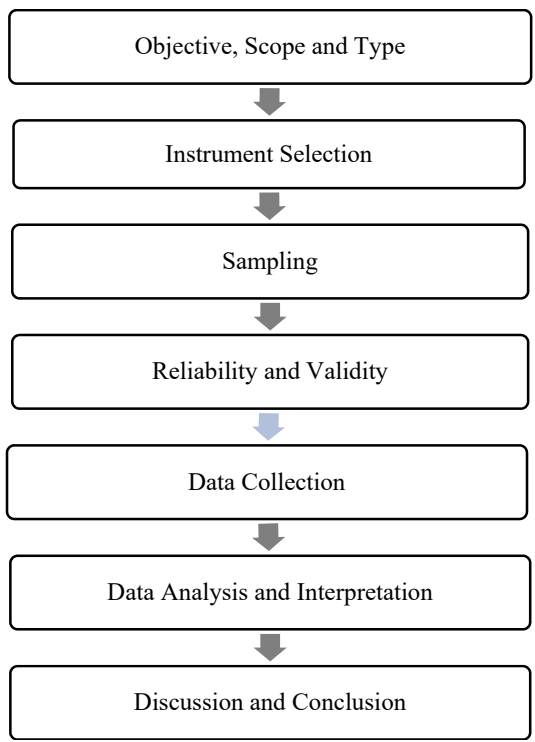

Figure 1 - Research design

**Instrument Selection**

This is a critical step, as it maps research problems to a mathematical domain. We asked students for the probability that they would choose testing careers by offering multiple choices: "Certainly Yes," "Yes," "Maybe," "No," and "Certainly Not." We have not found any prior study in this area. So, we asked the students to provide open-ended pros and cons and rationale in support of their decisions. We also asked the working professionals to give the pros and cons of a testing career.

**Sampling**

Our sample consisted of 73 undergraduate junior students of a computer-engineering program and 22 testing professionals. Out of 73, we had 70 valid responses with testing as a career choices as follows; 5 – "Yes," 10 – "Certainly Not," 22 – "No," and 33 – "Maybe." Missing or incomprehensible pros and cons rendered three responses invalid. We have tabulated the respondent details in Table 1. All 22 responses of the industry professionals were valid.

**Table 1: Chances of taking up testing career by junior computer engineering students**

| Response | Number |
|---|---|
| Certainly Yes | 0 |
| Yes | 5 |
| Maybe | 33 |
| No | 22 |
| Certainly not | 10 |
| **Total** | **70** |

While most of the students were admitted to the four year undergraduate engineering program after 12 years of schooling, a few of them (12) had lateral entries in the second year of the program after ten years of schooling followed by three years of an engineering diploma. The college is among the best in the state. It attracts bright students but has noticeable variation in performance in the entrance examinations and in the previous courses of the engineering program. Over half of the respondents were ambivalent to a testing career. That may imply other forces, such as the lack of positions as testers, at play. That is not true, especially for Indian markets where a job portal claims to have 15,401 open tester positions [18]. The test professionals were from various companies in a city and had a year or more of industry experience.

**Reliability and Validity**

A reliable and valid study contains the absence of bias and a higher truthfulness. The characteristics of a qualitative study are conceptualized as trustworthiness, rigor, and quality [19]. Lincoln and Guba [20] believe that in case of qualitative studies validity implies reliability and suggest a demonstration of only validity. Creswell and Miller [21] have observed qualitative researchers employing member checking, triangulation, peer reviews, thick description, and external audits to demonstrate validity. They have defined triangulation, a widely used measure in qualitative studies in this context, as a validity procedure where one searches for convergence among multiple and different sources of information to form themes or categories in a study. We asked the respondents to list pros and cons of testing careers and probability of them choosing a testing career along with rationale. We triangulated the rationale and pros-cons and found virtually no divergence between them.

**Data Collection**

We explained the background of our study to students and professionals in their respective sessions and sought their responses. We manually tagged all the responses and then iteratively coded them until no further code changes (merging or demerging) were possible.

Table 2 and Table 3 provide frequencies of various pros and cons from students with the choices listed above and aggregation of all choices.

Table 2: Frequencies of Pros of students, categorized by their testing career choice responses

| PRO -> Testing Career | Thinking Job | Learning Opportunities | Easy Life | Important Job | Higher salary | Growing Field | Interesting job |
|---|---|---|---|---|---|---|---|
| **Yes** | 5 | 2 | 1 | | | | |
| **Certainly Not** | 2 | 2 | 2 | 2 | | | |
| **No** | 10 | 8 | 2 | 2 | 1 | | |
| **May Be** | 16 | 14 | 3 | 8 | 1 | 3 | 3 |
| **Total** | 33 | 26 | 8 | 12 | 2 | 3 | 3 |
| **Percentage** | 47 % | 37 % | 11 % | 17 % | 3 % | 4 % | 4 % |

Table 3: Frequencies of Cons of students, categorized by their testing career choice responses

| CON -> Testing Career | 2nd class citizen | Miss development | Tedious | Difficult | Hard work | No patience | No interest | Other | Time pressure | Frustrating | Less thinking |
|---|---|---|---|---|---|---|---|---|---|---|---|
| **Yes** | 2 | 3 | | | | | | 2 | | | |
| **Certainly Not** | 4 | | 7 | | | | | | | | |
| **No** | 5 | 5 | 3 | 7 | 2 | 2 | 2 | 3 | | | |
| **May Be** | 14 | 7 | 9 | 1 | 1 | 3 | 4 | 2 | 5 | 4 | 4 |
| **Total** | 25 | 15 | 19 | 8 | 3 | 5 | 6 | 7 | 5 | 4 | 4 |

| Percentage | 36% | 21% | 27% | 11% | 4% | 7% | 9% | 10% | 7% | 6% | 6% |
|---|---|---|---|---|---|---|---|---|---|---|---|

The professionals' responses are tabulated in tables 4 and 5.

Table 4: Frequencies of Pros of professionals

| Thinking Job | Learning Opportunities | Important Job | Interesting job | Proximity to customer | Coding not required | In control | Growing Field |
|---|---|---|---|---|---|---|---|
| 13 | 12 | 11 | 4 | 2 | 2 | 2 | 1 |
| 59% | 55% | 50% | 18% | 9% | 9% | 9% | 5% |

Table 5: Frequencies of Cons of professionals

| 2nd class citizen | Stressful | Frustrating | Tedious | Miss development | Patience |
|---|---|---|---|---|---|
| 16 | 9 | 5 | 2 | 2 | 1 |
| 73% | 41% | 23% | 9% | 9% | 5% |

**Data Analysis and Interpretation**

The analysis of pros and cons resulted in the following key factors;

- Thinking Job – This encompasses views such as challenging, creative, innovative, and requiring logical and analytical thinking.

- Learning Opportunities – This primarily includes learning different technologies and techniques from students' perspectives, and the product life cycle, business processes, and the entire product from professionals' perspectives.

- Important Job – Students and professionals both realize that testers are gatekeepers and are accountable and responsible for the product quality. They, therefore, regard testing as an important job.

- Second-class citizen – This is a major factor and it is commonly voiced. It includes testers not being involved in decision making, testers being blamed for bad quality, the developers being rewarded for good quality, and testers getting lower pay, being perceived as less important, experiencing subdued career growth, etc.

- Tedious – This refers to the repetitive nature of testing.

- Miss development – This relates to testers not developing code or software.

- Stressful – This covers giving testers insufficient time yet still holding them responsible for product quality.

- Frustrating - This includes not finding defects, inability to reproduce defects, and having to deal with different versions and vendors of third party software.

**Students**

While nobody chose the "Certainly Yes" option, only five students chose the "Yes" option. All five perceived testing as a thinking job. Two of them felt that it offers better learning opportunities and only one considered it to be an easy job. The cons were that they will miss development, they will have a lower salary, they will grow in the field more slowly (will become second-class citizens), and their performance will depend on others. Two students each quoted these.

Ten students vehemently (by selecting "Certainly Not" option) refused to choose the testing career. They obviously did not see many benefits. Two students each quoted the following: thinking job, learning opportunities, important job, and easy job. On the con side, seven of them perceived testing to be a tedious job and four felt that they would be relegated to the status of second-class citizens. It seems that they had just made up their minds and had not thought very deeply on the issue.

Twenty-two students would not like to go for testing careers even though ten of them believed it to be a thinking job and eight of them agreed that it offered learning opportunities. Two students each felt it was an important and easy job. One believed that testers get higher salary. On the con side, seven of them felt it was a difficult job and five each sensed that they would miss development and become second-class citizens. Three felt it to be tedious and two each opined that the job would require more effort, more patience, and they would not be interested in it. The variety of reasons seems to have gone up.

Out of the 33 students who were unsure about embarking on the testing career, 16 perceived it to be a thinking job, 14 thought it offered learning opportunities, and eight sensed that this would be an important job. Three students each opined it to be a growing area, an easy area, and an interesting area. On the con side, a large number of students (14) expressed the fear of becoming second-class citizens. Nine felt it to be tedious and seven mentioned that they would miss development. Five felt that there would be severe deadline pressures and four students each felt that testing did not require thinking, it was a frustrating job, and it did not interest them. Three of them confessed that they did not have the required patience. Some of the students made ambivalent statements such as, "Ability to think increases" as a pro and "Does not help for innovation" as a con; "No Coding" as a Pro and "Missing development" as a Con; "Interesting Field" as a pro and "Boring Life" as con. Either they were unclear or looking at the situation from different perspectives.

It is also interesting to look at number of pros and cons listed by students and their choices. We have tabulated them below;

Table 6: The number of Pros and Cons of students based on their choices

| Choice | Number | Pro Reasons | Pro Ratio | Con Reasons | Con Ratio |
|---|---|---|---|---|---|
| Yes | 5 | 8 | 1.60 | 7 | 1.40 |
| May Be | 33 | 48 | 1.45 | 54 | 1.63 |
| No | 21 | 23 | 1.10 | 29 | 1.38 |
| Certainly Not | 9 | 8 | 0.89 | 11 | 1.22 |

The ratio of pros per student decreases as we move from "Yes" to "Certainly Not." The number of cons, on the other hand, is the highest for "May be" for students and the lowest for "Certainly Not." The former may have been thinking about the testing options and have come across many cons and the latter may have made up their mind and have not thought much about the cons. The con ratio for "Yes" and "No" is about the same. The former have a lower con ratio than pro, while for the latter, it is other way.

The top three pros and cons for all the students are given in Table 7.

Table 7: The top three Pros and Cons for all the students

| Pro | Count | Percentage | Con | Count | Percentage |
|---|---|---|---|---|---|
| Thinking Job | 33 | 47% | Second-Class Citizen | 25 | 36 % |
| Learning Opportunities | 26 | 37% | Tedious | 21 | 30 % |
| Important Job | 12 | 17% | Miss Development | 15 | 21 % |

The other notable pros pointed out by the students were "easy life" (11%), "growing field," (4%) and "interesting field" (4%). The cons were "more difficult work" (11%), "no interest" (9%), and "no patience" (7%) and "deadline pressure" (7%).

**Professionals**

We have tabulated the pros and cons provided by the top three industry professionals in Table 8 as follows;

Table 8: The top three Pros and Cons for the professionals

| Pro | Count | Percentage | Con | Count | Percentage |
|---|---|---|---|---|---|
| Thinking Job | 13 | 59% | Second-Class Citizen | 16 | 73% |
| Learning Opportunities | 12 | 55% | Stressful | 9 | 41% |
| Important Job | 11 | 50% | Frustrating | 5 | 23% |

The other notable pros pointed out by the test professionals were "interesting job" (11%), "proximity to customer," "being in control," and "coding not required" (9% each). The cons were "missing development" and "tedious" (9% each).

The similarity is striking. The top three pros and their ranking are the same in both groups. The industry professionals appear to be unanimous about the pros with almost equal voting for all three factors. Students have articulated learning opportunities as getting to know different products, technologies, and languages. The professionals, on the other hand, have included complete product knowledge and product life cycle. In case of the cons, there are differences among students and professionals. The professionals are very emphatic about being the second-class citizen and opine that there is more stress and frustration in testing careers. The students' division appears to be uniform across the three factors – "second class citizen," "tedious," and "miss development."

**Discussion**

It is evident that the testing profession is far from being popular. Less than seven percent of the students are thinking of taking up testing careers. While 46% of the students were ambivalent, the authors have seen them veering away from a testing career during placement. One reason for some of them to choose the testing profession is the opportunity to join branded organizations, such as Microsoft, Google, or Amazon.

This apathy towards testing careers has less to do with a perception and knowledge gap about the testing profession and more to do with the way the testing profession and the test professionals are treated in the industry. The students are very well aware of the challenges and opportunities for creativity that the testing profession offers. They also appreciate the learning opportunities that the profession offers. Their focus, in the case of learning, is more on tools and languages than on business problems, though. We can make them aware of this aspect with the help of invited lectures by industry stalwarts. While the developer is far away from the business customers and their problems, the tester enjoys their proximity, can learn immensely, and perhaps graduate easily into business analyst roles. Some students do appreciate the importance of testing activity and are aware that testers are responsible and accountable for the product quality. The percentage of such students is 17% as against the professionals 50%. We can apprise students of the complete product life cycle through real-life projects and exposure to industry processes.

The major issue is on the con side. The students are aware that the profession is relegated to second-class citizenship and vote that as the most critical issue. However, only 36% of the students believe so, as against 73% of the professionals. If students are exposed to this reality, many more may get distracted from the testing profession. The industry has to take care of this issue. While they may not have planned for this situation to occur, they have to plan to ensure that the situation does not occur. The professionals have cited examples such as not being involved in the decision process, not getting credit for good quality products but getting discredit for bad quality products, not having competitive growth paths, and exerting schedule pressure on testers to compensate for developers overruns. Some of these problems are relatively easy to fix and they must be fixed. We must add that the situation has been changing for the better over the years but not to the required levels.

Unlike the test professionals, some students believe testing is a tedious job and it robs them of professional development opportunity. This appears to be due to a lack of knowledge. Many organizations automate repetitive and regression testing activities. Test automation is actually a software development activity that uses scripting languages and environments similar to software product development. Further, organizations can have a different cadre such as "test technician" to carry out repetitive testing, wherever required. They also can have an intertwining career between testing and development as those skills complement each other.

**Conclusion**

Improvement in the quality of software systems will increase their usage and benefits. Software organizations are struggling with the challenge of software quality and must adopt a variety of approaches. They have to build the 'quality in' by working on all the phases including the last phase of testing. Further, they have to start working on the people aspects of software development activities. We have presented here a sliver of the people issue – why engineers do not choose testing careers.

We have found that very few engineering graduates have a testing career on their minds despite the challenges and learning requirements of the testing activity. They have reasonably good understanding – in terms of the positive and negative aspects – of the testing career and still do not want to take it up. In fact, if they develop better understanding with the help of testing professionals, likelihood of they taking up the testing career might diminish even further. The main reason is, the testing professionals believe that they are treated as second-class citizens i.e. they are not involved in the decision making process, and while they are reprimanded for failures they are not rewarded for the success of products they have tested.

We need to repeat the study in different geographical locations to validate the findings. The segmentation in testing business organizations, total testing experience, different colleges, different disciplines, and placement records may complete the picture. The impact of determinants such as gender, domicile, and academic performance may add more value. We also need to talk to business leaders and incorporate their views on the study findings.

We have suggested a few steps to bring in changes. Discussing them with a larger population to ascertain their acceptability and success will be helpful. In that sense, this is an initial study. While it provides some insights into the situation, it requires validation and reinforcement with wider experimentation.

**Acknowledgement:**

**References:**

1. Lions, J.L. *ARIANE 5 - Flight 501 Failure, report by the enquire board* Accessed Jan 2016; Available from: http://sunnyday.mit.edu/accidents/Ariane5accidentreport.html
2. Tan, G. *A Collection of Well-Known Software Failures*. 2012 28-Oct-2015]; Available from: http://www.cse.lehigh.edu/~gtan/bug/softwarebug.html.
3. NIST, *Summary of NIST Strategic Planning Results*. 2002, National Institute of Standards Technology.
4. Charette, R.N., *Why software fails.* IEEE spectrum, 2005. **42**(9): p. 36.
5. Micro-Focus, *Testing: IT's Invisible Giant*. 2011, Micro Focus.
6. Burnstein, I., G. Saxena, and R. Grom, *A testing maturity model for software test process assessment and improvement.* 1999.
7. Koomen, T. and M. Pol, *Test Process Improvement.* Essex: Pearson Education, 1999.
8. Kasurinen, J., O. Taipale, and K. Smolander, *Software test automation in practice: empirical observations.* Advances in Software Engineering, 2010. **2010**.
9. Graham, D. and M. Fewster, *Experiences of test automation: case studies of software test automation*. 2012: Addison-Wesley Professional.
10. Weyuker, E.J., et al., *Clearing a career path for software testers.* Software, IEEE, 2000. **17**(2): p. 76-82.
11. Broman, D., K. Sandahl, and M.A. Baker, *The company approach to software engineering project courses.* Education, IEEE Transactions on, 2012. **55**(4): p. 445-452.

12. Dagenais, B., et al. *Moving into a new software project landscape*. in *Proceedings of the 32nd ACM/IEEE International Conference on Software Engineering-Volume 1*. 2010. ACM.
13. Glass, R.L., I. Vessey, and V. Ramesh, *Research in software engineering: an analysis of the literature.* Information and Software technology, 2002. **44**(8): p. 491-506.
14. Rooksby, J., M. Rouncefield, and I. Sommerville, *Testing in the wild: The social and organisational dimensions of real world practice.* Computer Supported Cooperative Work (CSCW), 2009. **18**(5-6): p. 559-580.
15. Hazzan, O. and J. Tomayko, *Human aspects of software engineering: The case of extreme programming*, in *Extreme Programming and Agile Processes in Software Engineering*. 2004, Springer. p. 303-311.
16. Fernando Capretz, L., *Bringing the human factor to software engineering.* Software, IEEE, 2014. **31**(2): p. 104-104.
17. Shah, H. and M.J. Harrold. *Studying human and social aspects of testing in a service-based software company: case study*. in *Proceedings of the 2010 ICSE Workshop on Cooperative and Human Aspects of Software Engineering*. 2010. ACM.
18. Available from: http://www.shine.com/job-search/simple/software-testing/. Accessed on 24 Jan 2016;
19. Golafshani, N., *Understanding reliability and validity in qualitative research.* The qualitative report, 2003. **8**(4): p. 597-606.
20. Lincoln, Y.S. and E.G. Guba, *Naturalist inquiry.* Beverly Hills, CA: Sage, 1985.
21. Creswell, J.W. and D.L. Miller, *Determining validity in qualitative inquiry.* Theory into practice, 2000. **39**(3): p. 124-130.